\documentclass{webofc}
\usepackage[varg]{txfonts} 
\usepackage{hyperref}
\usepackage{url}
\hypersetup{colorlinks=true,citecolor=blue,urlcolor=blue,linkcolor=blue}
\usepackage{mathrsfs}
\usepackage{braket}
\newcommand{\vecr}{\mathbf{r}}
\newcommand{\teps}{\tilde{\epsilon}}

\begin{document}
\title{Collective-subspace requantization for sub-barrier fusion reactions: Inertial functions for 
collective motions}
%
%

\author{
\firstname{Chisato} 
\lastname{Ruike}\inst{1}
\and
\firstname{Kai} \lastname{Wen}\inst{2,3}
\and
\firstname{Nobuo} 
\lastname{Hinohara}\inst{3,4,5}
\and
\firstname{Takashi} 
\lastname{Nakatsukasa}\inst{3,4,6}
}

\institute{
Graduate School of Science and Technology, University of Tsukuba, Tsukuba, 305-8571, Japan
\and
KEK Theory Center, Institute of Particle and Nuclear Studies, High Energy Accelerator Reserach Organization, Tsukuba 305-0801, Japan
\and
Center for Computational Sciences, University of Tsukuba, Tsukuba 305-8577, Japan
\and
Faculty of Pure and Applied Sciences, University of Tsukuba, Tsukuba 305-8571, Japan
\and
Facility for Rare Isotope Beams, Michigan State University, East Lansing, Michigan 48824, USA
\and
RIKEN Nishina Center, Wako 351-0198, Japan
}

\abstract{%
The adiabatic self-consistent collective coordinate
(ASCC) method is used
to determine the optimum reaction path and
to calculate the potential and the inertial functions
of the reaction model.
The properties of the inertial functions are investigated
with the ASCC method,
in comparison with those of the cranking formulae.
In addition, the properties of the pair rotation are investigated
in the BCS pair model.
The moments of inertia for rotation in
both the real and the gauge spaces
may decrease as the deformation develops.

}
\maketitle
\section{Introduction \label{sec:intro}}

Time-dependent density functional theory (TDDFT) is a standard tool
for studies of a variety of heavy-ion reactions
\cite{Neg82,Sim12,NMMY16}, such as
nuclear fusion above the Coulomb barrier energy
\cite{GN12-P,Was15,EN15-P},
induced fission reaction \cite{BMRS16},
quasi-fission reaction \cite{UOS16},
and multi-nucleon transfer reactions \cite{Sim10,SY16}.
It also provides an important method in studies of
nuclear response,
namely the (quasiparticle) random-phase approximation ((Q)RPA)
in the linear regime.
In addition, the properties of the giant resonances
have been extensively
investigated with TDDFT real-time simulations
\cite{NY05,Eba10,SBMR11,SL13-2}.
A universal description of nuclei in the entire mass region and
a unified microscopic description of nuclear structure and reaction
are great features of the TDDFT method.

On the other hand,
the current TDDFT models of nuclei
have a problem in a proper account of fluctuation effects.
The problems are especially serious in transitional regions
where the nuclear shape fluctuates
and is not well-defined.
In nuclear reaction studies,
a prominent weak point is evident for
the sub-barrier fusion and the spontaneous fission reactions,
which never takes place in the straightforward TDDFT calculations.
The theory lacks a part of important quantum fluctuation
associated with the nuclear shape.
In order to recover the missing quantum fluctuation,
theories beyond the mean field are required.

The most popular method of improving the mean-field theory is
the generator coordinate method (GCM) \cite{RS80,Sch19}.
The GCM is a quantum mechanical approach
and is capable of describing the quantum fluctuations
along the chosen generator coordinates.
Increasing the number of generator coordinates,
in principle, the solution converges to the exact eigenstates
of the Hamiltonian.
However, this is not the case for the energy density functionals.
Although many of the nuclear energy density functionals are based on
density-dependent effective interactions,
the exact ground state of the Hamiltonian is not physical.
Therefore,
we must carefully choose a small number of generator coordinates
to avoid such unphysical states.
This also prevents us from utilizing a variational method
to determine the generator coordinates \cite{SONY06,Fuk13}.
The GCM has both numerical and theoretical problems \cite{NMMY16},
especially when we use the effective interactions
which depend on a fractional power of density
\cite{AER01,DSNR07,Dug09}.

Our strategy for a solution to these problems 
is the requantization of collective degrees of freedom
which are selected according to the TDDFT dynamics.
Similar approaches have been adopted in studies of
low-lying excitation spectra and in the fission dynamics,
e.g., the five-dimensional collective Hamiltonian
(5DCH) method instead of the GCM on the $(\beta,\gamma)$ plane
\cite{KB67,Del10,NMMY16}
and the collective model of spontaneous fission \cite{Sad20}.
However, in these models, the shape variables are chosen by hand
and the inertia parameters are approximated by
the cranking formula \cite{Ing54,Ker61}.
We aim to improve the models in terms of the microscopic point
of view.

We determine a collective subspace
which is decoupled from the other intrinsic
degrees of freedom.
For this purpose, we use the adiabatic self-consistent
collective coordinate (ASCC) method
\cite{MNM00,HNMM07,HNMM09,Nak12,NMMY16}.
The solution of the ASCC equations provides
a set of canonical variables $(q,p)$ to specify
the one-body density $\hat{\rho}(q,p)$, and
the generators of collective variables $(\hat{Q}(q),\hat{P}(q))$.
The collective potential is given by the energy values at
$p=0$, $V(q)=E[\rho(q,p=0)]$,
and the expansion with respect to the $p$ provides
the kinetic energy with the inertial masses,
$M^{-1}(q)=\partial^2 E/\partial p^2|_{p=0}$.
This leads to the collective Hamiltonian
\begin{equation}
H=\frac{p^2}{2M(q)} + V(q) .
\label{eq:collective_H}
\end{equation}
Since the scale of the coordinate $q$ is arbitrary,
we can adjust it so as to make the collective inertia constant,
$M(q)=\mu$.
In the study of the sub-barrier nuclear fusion,
we expect that the coordinate $q$ should be proportional to
the distance between the projectile and the target nuclei, $r$,
when two nuclei are far apart.
Thus, it is intuitive to write down the collective Hamiltonian
in the relative coordinate $r$ and its conjugate momentum
$p_r=p(dq/dr)$.
It should be noted that the collective inertia $M(r)$,
given by
$\{M(r)\}^{-1}=(dr/dq)^2 \mu^{-1}$,
is no longer constant,
especially at $r\lesssim R_{\rm touch}$
where $R_{\rm touch}$ is a touching distance of two nuclei.

Assuming the axial symmetry with respect to the axis of
the relative coordinate $\mathbf{r}$
between the projectile and the target,
two angles to fix the direction of $\mathbf{r}$
describe the rotational motion, and
their moments of inertia are calculated
solving the Thouless-Valatin equation \cite{TV62}
at each point on the collective subspace $r$.
Eventually, the following collective Hamiltonian is obtained.
\begin{equation}
H=\frac{p_r^2}{2M(r)} + \frac{L^2}{2\mathscr{I}(r)} + V(r) ,
\label{eq:collective_Hamiltonian}
\end{equation}
where $L^2$ is the square of the angular momentum
produced by the rotation of the coordinate $\mathbf{r}$.
In the end, the canonical quantization is performed,
replacing $p_r$ by $-i\partial/\partial q$ ($\hbar=1$)
with the Pauli's prescription
and $L^2$ by the angular momentum operator $\hat{L}^2$.
It should be noted that the point transformation from $q$ to $r$
does not change the physics, as far as the inertial functions
and the conjugate momentum are properly defined.

In this paper, we focus our discussion on the inertial functions
for various collective motions.
In Sec.~\ref{sec:collective_subspace}, we briefly review
the ASCC method and present methods of
calculating the inertial functions
in Sec.~\ref{sec:inertial_functions}.
We especially, describe the derivation of the cranking formulae
based on the TDDFT equation of motion,
which can clarify the missing correlations in the cranking formulae.
In Sec.~\ref{sec:pair_rotation},
we show a somewhat surprising property of
the moment of inertia for the pair rotation.
The conclusion and the perspectives are
given in Sec.~\ref{sec:conclusion}.

\newpage

\section{Collective subspace \label{sec:collective_subspace}}

Using the energy density functional $E[\rho]$
and the local (normal-mode) generator $Q(q)$,
the collective subspace for the low-energy collective motion
is defined by
\begin{equation}
\biggr[ h[\rho(q)] - \lambda Q(q), \rho(q) \biggr]= 0 ,
\label{eq:MF-MF}
\end{equation}
where $h_{ij}\equiv \delta E/\delta\rho_{ji}$.
The Lagrange multiplier $\lambda$
corresponds to the gradient of the energy,
$\lambda=\partial E[\rho(q)]/\partial q$.
The generators, $(Q(q),P(q))$, are given by a solution of
the moving-frame random-phase approximation (MF-RPA).
\begin{equation}
\left\{ A(q) + B(q) \right\}
\left\{ A(q) - B(q) \right\} Q(q) = \omega^2 Q(q),
\quad\quad P(q) = M(q)
\left\{ A(q) - B(q) \right\} Q(q) ,
\label{eq:MF-RPA}
\end{equation}
where the matrices $A(q)$ and $B(q)$ are given by
\begin{equation}
A_{php'h'}(q) = (\epsilon_p-\epsilon_h)\delta_{pp'}\delta_{hh'}
+\frac{\partial h_{ph}}{\partial\rho_{p'h'}} ,
\quad\quad
B_{php'h'}(q) = \frac{\partial h_{ph}}{\partial\rho_{h'p'}} .
\label{eq:MF-RPA-matrix}
\end{equation}
The MF-RPA equation determines only the particle-hole and
hole-particle matrix elements
of the generators, $(Q(q),P(q))$.
In this paper, we assume that their particle-particle and
hole-hole matrix elements vanish.
The single-particle energies $\epsilon_i$ in Eq.~(\ref{eq:MF-RPA-matrix})
are the eigenvalues of $h-\lambda Q(q)$.
The derivatives in Eq.~(\ref{eq:MF-RPA-matrix})
are calculated according to the finite-amplitude method (FAM)
\cite{NIY07,AN11,Sto11},
especially using the m-FAM scheme in Ref.~\cite{AN13}.

The MF-RPA equation (\ref{eq:MF-MF}) cannot fix
the magnitude of $Q(q)$.
Since these equations are scale invariant concerning
$Q(q)\rightarrow \alpha Q(q)$,
$P(q)\rightarrow \alpha^{-1}P(q)$,
and $M(q)\rightarrow \alpha^{-2} M(q)$,
we can arbitrarily normalize $Q(q)$.
In actual calculations, we set $M(q)$ a constant value, $M(q)=\mu$,
to determine the normalization of $Q(q)$
according to the weakly canonicity condition,
$
\mathrm{Tr}\left(\left[Q(q),P(q)\right] \rho(q) \right)=i
$.

The ASCC equations consist of Eqs.~(\ref{eq:MF-MF}) and
(\ref{eq:MF-RPA}) which should be solved self-consistently.
Among solutions of Eq.~(\ref{eq:MF-RPA}), we select
a normal mode of the lowest energy keeping the axial symmetry.
Note that $\omega^2$ can be negative.
The numerical computation produces a series of density
$\{ \rho(q_0), \rho(q_1),\cdots \}$ on the discretized
collective coordinate
$q_n = q_0+n\times \Delta q$.
This defines the collective subspace,
namely the reaction path in the present study.
The potential in Eq. (\ref{eq:collective_H}),
$V(q)=E[\rho(q)]$,
then, transformed into $V(r)=V(q(r))$
in Eq. (\ref{eq:collective_Hamiltonian}).

\section{Inertial functions \label{sec:inertial_functions}}

In this paper, we focus our discussion on the properties of
collective inertias and their effect on the nuclear dynamics.

\subsection{ASCC inertial functions \label{sec:ASCC_mass}}

Since the ASCC method provides a decoupled collective subspace,
the inertial mass tensor should be block-diagonal:
no off-diagonal elements between
the collective subspace and the rest.
The value of the constant mass $\mu$
determines the scale of the coordinate $q$,
then,
the inertial function $M(r)$ in Eq. (\ref{eq:collective_Hamiltonian})
is given by
\begin{equation}
M(r)=\mu \left(\frac{dq}{dr}\right)^2 ,
\quad\quad
\left(
M_{ij}(r)= \sum_{kl}
\frac{\partial q^k}{\partial r_i}
\mu_{kl}
\frac{\partial q^l}{\partial r_j}
\right)
.
\label{eq:ASCC_mass}
\end{equation}
The second equation in the bracket corresponds to the case that
the collective subspace is multi-dimensional, in general.

The rotational modes correspond to the zero-frequency solutions of
Eq. (\ref{eq:MF-RPA}).
For the axially symmetric states, we should obtain
two independent solutions corresponding to the rotation around
the axes perpendicular to the symmetry axis.
Since these modes violate the axial symmetry ($K\neq 0$),
they are exactly decoupled from
the adopted axially symmetric mode ($K=0$),
$(Q(q),P(q))$
in Sec.~\ref{sec:collective_subspace}.

Although the rotational moments of inertia can be deduced
from the zero-frequency solutions of Eq. (\ref{eq:MF-RPA}),
they can be even more easily obtained from
the calculation of the strength function
at the zero frequency \cite{Hin15}.
We use this method to calculate the moments of inertia
$\mathscr{I}(r)$ in Eq. (\ref{eq:collective_Hamiltonian}).
This is equivalent to solutions of the Thouless-Valatin equations
\cite{TV62,RS80} with the moving-frame Hamiltonian.

\subsection{Cranking mass \label{sec:cranking_mass}}

The most popular method to estimate the collective masses
is the cranking formula \cite{Ing54,Ker61}.
It has been utilized in a variety of applications.
However, it has been known for many years that the collective masses
of the cranking formula have serious problems
with missing dynamical residual effects.

\subsubsection{Translational motion \label{sec:translational_motion}}

In this section, first, we present a derivation of
the cranking formula
for the translational motion, and demonstrate the importance of
the residual effects.

In the TDDFT,
a nucleus in the ground state is given by the stationary
condition for the density $\rho$.
\begin{equation}
\label{eq:stationary_equation}
\left[ h_0, \rho_0 \right] = 0 ,
\end{equation}
where $\rho_0$ is the one-body density of
the ground state.
$h_0$ is the Kohn-Sham (KS) Hamiltonian at the density $\rho_0$
which is represented by the KS orbitals $\ket{\phi_i}$,
defined by the KS equation
$h_0\ket{\phi_i}=\epsilon_i\ket{\phi_i}$,
as
\begin{equation}
\rho_0=\sum_{h=1}^A \ket{\phi_h} \bra{\phi_h} .
\end{equation}
The center of mass of $\rho_0$ is arbitrary and
chosen at the origin.
Let us consider a nucleus in the ground state,
moving in the $x$ direction with a constant velocity $v$.
The density is given as
\begin{equation}
\rho= e^{imv\hat{x}} e^{-ix(t)\hat{p}}\rho_0
e^{ix(t)\hat{p}} e^{-imv\hat{x}} ,
\label{eq:density_of_moving_nucleus}
\end{equation}
where $m$ is the nucleon mass.
The expectation values of the center of mass $X$
and the total momentum $P$ are
\begin{eqnarray}
\label{eq:CoM}
&&X=\frac{1}{A}\mathrm{Tr}\left(\hat{x}\rho\right)
= \frac{1}{A} \mathrm{Tr}\left(\hat{x}\rho_0\right)
+ \frac{x(t)}{A} \mathrm{Tr}\left(\rho_0\right) =x(t) ,\\
\label{eq:Total_momentum}
&&P=\mathrm{Tr}\left(\hat{p}\rho\right)
= \mathrm{Tr}\left(\hat{p}\rho_0\right)
+mv \mathrm{Tr}\left(\rho_0\right) =Amv .
\end{eqnarray}
Here, we use the commutation relation $[\hat{x},\hat{p}]=i$
and the fact that the trace of a product is commutable.
Equation (\ref{eq:Total_momentum})
indicates that the mass for the translational
motion is nothing but the total mass $Am$.
These equations, (\ref{eq:CoM}) and (\ref{eq:Total_momentum}),
are valid at any values of $x$ and $v=\dot{x}$.
Assuming $x$ and $v$ are small, we may linearize the density
variation as
$\delta\rho=x\delta\rho^{(x)}+v\delta\rho^{(v)}$
with
$\delta\rho^{(x)}=-i[\hat{p},\rho_0]$ and
$\delta\rho^{(v)}=im[\hat{x},\rho_0]$.
It is easy to verify that $\rho=\rho_0+\delta\rho$ produces
the same results as
Eqs. (\ref{eq:CoM}) and (\ref{eq:Total_momentum}).
Note that the particle-particle and hole-hole matrix elements
of $\delta\rho$ vanish,
$\bra{\phi_p}\delta\rho\ket{\phi_{p'}}
=\bra{\phi_h}\delta\rho\ket{\phi_{h'}}=0$.

The density of
Eq. (\ref{eq:density_of_moving_nucleus})
should satisfy the TDDFT equation
$i\dot{\rho}=\left[ h[\rho], \rho \right]$.
Now, we linearize the equation as
\begin{equation}
i\frac{\partial}{\partial t} \left(
x\delta\rho^{(x)} + v\delta\rho^{(v)} \right)
=\left[ h_0, x\delta\rho^{(x)}+v\delta\rho^{(v)} \right]
+ \left[ x\delta h^{(x)}+v\delta h^{(v)},\rho_0\right] ,
\label{eq:Linearized_TDDFT}
\end{equation}
where $\delta h=x\delta h^{(x)}+ v\delta h^{(v)}$
is the residual field induced by
the density variation,
$\delta h^{(i)}
= \partial h/\partial \rho |_{\rho_0}\cdot \delta\rho^{(i)}$
with $i=x, v$.
The term proportional to $x$, $x\delta h^{(x)}$,
represents a change in the KS potential
for a shift of the center of mass,
that is present when the translational symmetry is violated
in $h_0$.
The term proportional to $v$, $v\delta h^{(v)}$,
is induced by the velocity of the nucleus,
which appears when the KS potential has the velocity dependence.
The left-hand side of Eq. (\ref{eq:Linearized_TDDFT}) is
\begin{equation}
\mathrm{LHS}=
i\frac{\partial}{\partial t} \left(
x\delta\rho^{(x)} + v\delta\rho^{(v)} \right)
= iv \delta\rho^{(x)}
=v\left[\hat{p},\rho_0 \right] ,
\label{eq:TDDFT_lhs}
\end{equation}
and the right-hand side is
\begin{equation}
\mathrm{RHS}=
\left[ h_0, \delta\rho \right]
+ \left[ \delta h,\rho_0\right]
=
v \left[ h_0, \delta\rho^{(v)} \right]
+ v \left[ \delta h^{(v)},\rho_0\right] .
\label{eq:TDDFT_rhs}
\end{equation}
Here, we use the condition that the density
(\ref{eq:density_of_moving_nucleus})
at any center-of-mass coordinate with $v=0$
must satisfy the stationary equation (\ref{eq:stationary_equation}).
The particle-hole and hole-particle matrix elements
of Eq. (\ref{eq:Linearized_TDDFT}) lead to
\begin{eqnarray}
&&\bra{\phi_p}\hat{p}\ket{\phi_h}=
\left( \epsilon_p-\epsilon_h\right)
\bra{\phi_p}\delta\rho^{(v)}\ket{\phi_h}
+\bra{\phi_p}\delta h^{(v)}\ket{\phi_h}
, \\
&&\bra{\phi_h}\hat{p}\ket{\phi_p}=
\left( \epsilon_p-\epsilon_h\right)
\bra{\phi_h}\delta\rho^{(v)}\ket{\phi_p}
+\bra{\phi_h}\delta h^{(v)}\ket{\phi_p} .
\end{eqnarray}
The expectation value of the total momentum is calculated as
\begin{equation}
P=\mathrm{Tr}\left(\hat{p}\delta\rho\right)
= v\mathrm{Tr}\left(\hat{p}\delta\rho^{(v)}\right)
= v \sum_{p,h} \frac{\bra{\phi_p}\hat{p}\ket{\phi_h}
\bra{\phi_h}\left(\hat{p}-\delta h^{(v)}\right)\ket{\phi_p}
+\mathrm{c.c.}}{\epsilon_p-\epsilon_h} ,
\label{eq:Total_momentum_2}
\end{equation}
which should be equal to $Amv$,
as we have shown in Eq. (\ref{eq:Total_momentum}).

The cranking formula is obtained by the approximation
to neglect the residual field $\delta h^{(v)}$ in
Eq. (\ref{eq:Total_momentum_2}), that results in
the mass of the translational motion,
\begin{equation}
M= \frac{dP}{dv} = 2 \sum_{p,h} 
\frac{\left|\bra{\phi_p}\hat{p}\ket{\phi_h}\right|^2}
{\epsilon_p-\epsilon_h} .
\label{eq:cranking_formula_translation}
\end{equation}
In general, the cranking formula
(\ref{eq:cranking_formula_translation})
produces $M\neq Am$.
This is due to the violation of the Galilean invariance
of the static KS Hamiltonian $h_0$.
In fact, if $h_0$ satisfies the relation,
\begin{equation}
i \left[ h_0, \hat{x} \right] = \frac{\hat{p}}{m} ,
\label{eq:Galilean_invariance}
\end{equation}
then, we can rewrite the quantity
\begin{equation}
\left[ \hat{p}, \rho_0 \right]
= im \left[ \left[ h_0,\hat{x} \right], \rho_0 \right]
= im \left[ h_0, \left[ \hat{x},\rho_0 \right] \right]
= \left[ h_0, \delta\rho^{(v)} \right] ,
\end{equation}
where we use Eq. (\ref{eq:stationary_equation}).
This guarantees that the second term on the right-hand side of
the TDDFT equation (\ref{eq:TDDFT_rhs}) vanishes.
See Eqs. (\ref{eq:TDDFT_lhs}) and (\ref{eq:TDDFT_rhs}).
Thus, the cranking formula gives the exact total mass, $M=Am$.
In other words,
when Eq. (\ref{eq:Galilean_invariance}) is violated for $h_0$,
the residual field $\delta h^{(v)}$ is necessary
to restore the Galilean invariance
of the energy density functional $E[\rho]$.

\subsubsection{Rotational motion \label{sec:rotational_motion}}

The arguments for the translational motion can be
directly applicable to the motion of the uniform rotation.
When the ground-state solution of the KS equation
violates the rotational symmetry,
the rotational degrees of freedom emerge
from the orientation of the deformed intrinsic state.
Assuming the rotation around the $x$ axis with a constant
angular velocity $\omega=\dot{\theta}$,
the density is given,
similar to Eq. (\ref{eq:density_of_moving_nucleus}),
by
\begin{equation}
\rho= e^{iI\omega\hat{\theta}}
e^{-i\theta(t)\hat{j}_x} \rho_0
e^{i\theta(t)\hat{j}_x} e^{-iI\omega\hat{\theta}} ,
\label{eq:density_of_rotating_nucleus}
\end{equation}
where the angle operator $\hat{\theta}$
is a conjugate variable to the angular momentum $\hat{j}_x$,
namely $[\hat{\theta},\hat{j}_x]=i$.
$I=\mathscr{I}_x/A$ is linked to the moment of inertia
of the nucleus, $\mathscr{I}_x$.
Here, we only assume the existence of the angle operator
$\hat{\theta}$ and the moment of inertia $\mathscr{I}$,
but do not need the exact expression of those quantities\footnote{
This is different from the case of translational motion.
We explicitly know both the center-of-mass coordinate,
the total linear momentum, and the exact value of
the total mass in Sec.~\ref{sec:translational_motion}.
}.
In addition, in contrast to the translational case
in Sec.~\ref{sec:translational_motion},
the decoupling between the rotational
and the intrinsic motions is not exact.
Therefore, strictly speaking,
Eq. (\ref{eq:density_of_rotating_nucleus}) is true
only when $\omega=0$,
and the density $\rho_0$
in Eq. (\ref{eq:density_of_rotating_nucleus})
should change at a non-zero value of $\omega$.
In this section, we assume a small value of $\omega$
and neglect the $\omega$-dependence of the intrinsic structure.

The rest of the derivation is the same as those in
Sec.~\ref{sec:translational_motion}.
Neglecting the residual fields, $\delta h^{(\omega)}$,
the cranking formula for the moment of inertia is given by
\begin{equation}
\mathscr{I}_x= \frac{dJ}{d\omega}
= 2 \sum_{p,h} \frac{\left|\bra{\phi_p}\hat{j}_x\ket{\phi_h}\right|^2}
{\epsilon_p-\epsilon_h} .
\label{eq:cranking_formula_rotation}
\end{equation}

In the rotational case, there is no exact relation analogous to
the Galilean invariance.
Nevertheless, the condition in analogy to
Eq.~(\ref{eq:Galilean_invariance}) results in
the justification of the neglect of the residual effect.
Therefore, the cranking formula
(\ref{eq:cranking_formula_translation})
is valid (only) when
the KS potential has no velocity dependence.
In general, the residual terms associated
with $\delta h^{(\omega)}$ proportional to $\omega$
(terms of $\delta h^{(v)}$ in Eq.~(\ref{eq:Total_momentum_2}))
is necessary to give a proper value for the moment of inertia.

\subsubsection{Collective motion on the reaction path \label{sec:collective_motion}}

Instead of determining the collective subspace
by solving Eq. (\ref{eq:MF-MF}),
the generator $\hat{Q}(q)$ is most commonly replaced
by a constraining one-body operator given by hand, $\hat{C}$.
The constrained KS density is obtained by
\begin{equation}
\left[ h_\lambda, \rho_\lambda
\right] = 0 , \quad\quad
h_\lambda \equiv h[\rho_\lambda] - \lambda \hat{C}
\end{equation}
where $\lambda$ is the Lagrange multiplier.
We may change the label $\lambda$ into
the collective coordinate, for instance,
the distance between two colliding nuclei $r$.
The relation between $\lambda$ and $R$ is
given by $r=\mathrm{Tr}(\hat{r}\rho_\lambda)$.
The choice of the operator $\hat{r}$ may not be unique,
but a possible choice is given in Ref.~\cite{WN22}.
The density is represented by
the KS orbitals $\ket{\phi_i(r)}$ which are solutions of
$
h_\lambda\ket{\phi_i(r)}=\epsilon_i(r)\ket{\phi_i(r)}
$.

The cranking formula of
the inertial mass for the collective motion
with respect to the coordinate $r$ can be given,
in analogy to Eqs. (\ref{eq:cranking_formula_translation})
and (\ref{eq:cranking_formula_rotation}),
as
\begin{equation}
M^{\rm np}(r)
= 2 \sum_{p,h}
\frac{|\bra{\phi_p(r)} \partial/\partial r
	\ket{\phi_h(r)}|^2}{\epsilon_p(r)-\epsilon_h(r)} .
\label{eq:NP_cranking}
\end{equation}

The formula (\ref{eq:NP_cranking}) is often called
``non-perturbative cranking formula.''
This is why we put the superscript for $M^{\rm np}$.
A further approximation can give
``perturbative cranking formula'' \cite{WN22},
denoted as $M^{\rm p}$ in this paper,
\begin{equation}
M^{\rm p}=\frac{1}{2} S_3(\hat{C},\hat{C})
\left\{ S_1(\hat{r},\hat{C}) \right\}^{-2} ,
\label{eq:P_cranking}
\end{equation}
with
\begin{equation}
S_k (\hat{A},\hat{B})\equiv
\frac{1}{2}
\sum_{p,h}
\frac{\bra{\phi_h}\hat{A}\ket{\phi_p}
\bra{\phi_p}\hat{B}\ket{\phi_h}+\mbox{c.c.}}
{\left(\epsilon_{p}-\epsilon_h\right)^k}
=S_k(\hat{B},\hat{A}) ,
\label{S_k}
\end{equation}
where $\hat{A}$ and $\hat{B}$ are the one-body operators.
In the case of $\hat{A}=\hat{B}$,
$S_k(\hat{A},\hat{A})$
corresponds to the $k$th inverse energy-weighted sum-rule value,
$m_k(\hat{A})\equiv S_k(\hat{A},\hat{A})$.
In the case of $\hat{C}=\hat{r}$,
the perturbative cranking formula has become an even simpler form
\begin{equation}
M^{\rm p}=\frac{1}{2} m_3(\hat{r})
\left\{ m_1(\hat{r}) \right\}^{-2} ,
\end{equation}
which has been extensively utilized in the literature.

\section{Results \label{sec:results}}

\subsection{Inertial functions
for sub-barrier nuclear reactions}

The energy density functional
is slightly modified from $E_{\rm BKN}[\rho]$
of Ref.~\cite{BKN76}
to allow the KS potential to be velocity-dependent.
\begin{equation}
E[\rho]=E_{\rm BKN}[\rho]
+B_3 \int d\vecr \left\{ \rho(\vecr)\tau(\vecr)-\mathbf{j}^2(\vecr)\right\} ,
\end{equation}
where $\rho(\vecr)$ is the isoscalar density, $\tau(\vecr)$ is the isoscalar kinetic density,
and $\mathbf{j}(\vecr)$ is the isoscalar current density.
The positive value of the parameter $B_3$ produces
the effective mass smaller than the bare nucleon mass,
$m^*/m < 1$.

Most of the numerical results are shown in the paper \cite{WN22}.
Here, we recapitulate the summary of the results.
The ASCC inertial functions calculated
according to Eq. (\ref{eq:ASCC_mass})
reproduce
\begin{enumerate}
\vspace{-5pt}
\item the total mass $Am$ for the translation,
\vspace{-5pt}
\item the reduced mass, $\mu_{\rm red}=A_1 A_2 m/(A_1+A_2)$,
for the relative motion
at large $r$ ($r \gg R_{\rm touch}$),\vspace{-5pt}
\item the reduced value of the moment of inertia,
$\mathscr{I}_{\rm red}\equiv \mu_{\rm red}r^2$ , at large $r$.
\vspace{-5pt}
\end{enumerate}
On the other hand,
the cranking formulae, Eqs.~(\ref{eq:cranking_formula_translation}),
(\ref{eq:cranking_formula_rotation}),
(\ref{eq:NP_cranking}),
and (\ref{eq:P_cranking}),
fail to reproduce these quantities.
Instead, the cranking formulae approximately produce
\vspace{-5pt}
\begin{enumerate}
\item the total effective mass,
$A\langle m^* \rangle$
for the translational motion,
\vspace{-5pt}
\item the reduced effective mass
$\mu^*_{\rm red}=A_1 A_2 \langle m^* \rangle /(A_1+A_2)$
at large $r$
\vspace{-5pt}
\item the moments of inertia
$\mu^*_{\rm red}r^2$ at large $r$
\end{enumerate}
\vspace{-5pt}
These results apparently indicate the failure of the
cranking formulae and the superiority of
the ASCC masses over the cranking masses.

We have also found an interesting property
of the moments of inertia,
opposite to our naive intuition:
the calculated values
decrease in the vicinity of the ground state
as a function of the relative distance $r$,
or equivalently as a function of the nuclear deformation $\beta$.
\begin{equation}
\frac{d\mathscr{I}}{dr}<0 , \quad\quad
\frac{d\mathscr{I}}{d\beta}<0 .
\label{eq:dIdr<0}
\end{equation}
This is due to the property that
the moment of inertia is close to the rigid-body value
at the ground state (without pairing),
$\mathscr{I}\approx \mathscr{I}_{\rm rig}$,
then, immediately drops to the reduced value
$\mathscr{I}_{\rm red}$
once two nuclei are apart, $r\gtrsim R_{\rm touch}$.
This is a consequence of the quantum mechanical requirement.
The total moment of inertia for two separated nuclei
can be decomposed as
\begin{equation}
\mathscr{I}(r)=\mathscr{I}_1 + \mathscr{I}_2
+ \mathscr{I}_{\rm red}(r) ,
\end{equation}
where $\mathscr{I}_i$ ($i=1,2$)
is the moment of inertia of the nucleus $i$
concerning its own center of mass.
When the two nuclei are both spherical,
the quantum mechanics requires
$\mathscr{I}_1 = \mathscr{I}_2 = 0$
and produces the reduced value,
$\mathscr{I}(r)=\mathscr{I}_{\rm red}$
at $r\gtrsim R_{\rm touch}$.

For more details and possible effects on
the fusion cross sections,
the reader is referred to Ref.~\cite{WN22}.

\section{Moments of inertia for pair rotation \label{sec:pair_rotation}}

In a somewhat different context,
we have recently found a property, similar to
Eq. (\ref{eq:dIdr<0}),
that the moments of inertia
of the pair rotation decreases as a function of
the pairing gap (deformation in the gauge space) $\Delta$.
\begin{equation}
\frac{\partial\mathscr{I}_{\rm pair}}{\partial\Delta}<0 ,
\label{dIdDelta<0}
\end{equation}
where the moment of inertia is calculated,
using the ground-state energies with different neutron (proton)
numbers, as
\begin{equation}
\frac{1}{\mathscr{I}_{\rm pair}} = \frac{E(N+2)- 2E(N) + E(N-2)}{4} .
\end{equation}

This behavior can be qualitatively understood as follows,
using the BCS treatment of the pair model.
The particle number $N$ is written as
\begin{equation}
N(\lambda,\Delta)=\int_{\lambda-\Lambda}^{\lambda+\Lambda}
\rho(\epsilon) g(\epsilon-\lambda,\Delta)d\epsilon ,
\quad\quad
g(\teps,\Delta)\equiv\frac{1}{2}
\left( 1-\frac{\teps}{\sqrt{\teps^2+\Delta^2}} \right) ,
\end{equation}
where $\rho(\epsilon)$ is the single-particle level density and
$\lambda$ is the chemical potential.
We truncate the model space for the pairing correlations
to include the single-particle levels
in the section of $2\Lambda$ centered around the chemical potential.
The moment of inertia is given by
\begin{equation}
\mathscr{I}_{\rm pair}=\frac{\partial N}{\partial\lambda}
=\int_{\lambda-\Lambda}^{\lambda+\Lambda}
\rho(\epsilon) \frac{\partial g(\epsilon-\lambda,\Delta)}{\partial\lambda} 
d\epsilon
=-\int_{-\Lambda}^{\Lambda}
\rho(\teps+\lambda) \frac{\partial g(\teps,\Delta)}{\partial\teps} d\teps.
\label{eq:I}
\end{equation}
Here, we keep the model space invariant when we differentiate
the equation with respect to $\lambda$.
Then, its derivative with respect to $\Delta$ is
\begin{equation}
\frac{\partial\mathscr{I}_{\rm pair}}{\partial\Delta}=\frac{\partial^2 
N}{\partial\lambda\partial\Delta}
=-\int_{-\Lambda}^{\Lambda}
d\teps \rho(\teps+\lambda) \frac{\partial^2 
g(\teps,\Delta)}{\partial\teps\partial\Delta}
= -\frac{1}{2\Delta}\int_{-\Lambda/\Delta}^{\Lambda/\Delta} 
\rho(\lambda+x\Delta)
\frac{1-2x^2}{(x^2+1)^{5/2}} dx .
\label{eq:dIdDelta}
\end{equation}
In the energy interval of $2\Lambda$,
we approximate the level density in the linear order as
\begin{equation}
\rho(\lambda+x\Delta) \approx \rho(\lambda) + x\Delta\rho'(\lambda) ,
\end{equation}
which is inserted into Eq. (\ref{eq:dIdDelta}) to obtain
\begin{equation}
\frac{\partial\mathscr{I}_{\rm pair}}{\partial\Delta}=
-\frac{\rho(\lambda)}{\Delta} I(\Lambda/\Delta) ,
\end{equation}
and
\begin{equation}
I(\eta) \equiv\int_{0}^{\eta} dx \frac{1-2x^2}{(x^2+1)^{5/2}}
=\frac{\eta}{(\eta^2+1)^{3/2}} .
\end{equation}
For $\eta=\Lambda/\Delta>0$, it is trivial to have $I(\eta)>0$.
Thus, Eq. (\ref{dIdDelta<0}) is proved.

\section{Conclusion and future perspectives \label{sec:conclusion}}

Using the ASCC method, we have microscopically derived
a collective subspace (reaction path)
for the sub-barrier fusion reaction.
In addition, the method provides the potential and
the inertial functions,
which leads to the collective Hamiltonian of the reaction model.
In this paper, we particularly focus our discussion on
properties of the inertial functions.

The ASCC inertial functions perfectly agree with the correct
values in the asymptotic region, such as the total mass for
the translational motion, the reduced mass for the relative motion
at large $r$, and the reduced moments of inertia
at large $r$.
On the other hand, the cranking formulae fail to reproduce
these asymptotic values.
This indicates the superiority of the ASCC method
over the cranking formula.

The moment of inertia at the ground state is approximately
equal to that of the rigid-body value,
but surprisingly, it decreases as the deformation increases.
We may expect that the isotropic velocity distribution
at the equilibrium is distorted to be anisotropic,
leading to a transition from the ``rigid body'' to
the ``irrotational flow.''
Since the moment of inertia for the irrotational fluid
vanishes for spherical systems,
the two spherical nuclei, whose mass numbers $A_1$ and $A_2$,
at distance $r$, produce the moment of inertia, $\mu_{\rm red}r^2$.
This is well reproduced by ASCC moments of inertia,
but not for the cranking formula.

Further studies on the collective masses are desired,
particularly effects of pairing and beyond the mean field.
We have shown that the recovery of the Galilean invariance
violated in the KS potential is essential to produce
the collective inertia.
Since the pair field is known to break the Galilean
invariance \cite{BM75}, it is of great interest to
investigate residual effects of the pair fields.

We have also studied properties of the pair rotation
in the gauge space,
then, have found that the moment of inertia
also decreases as the pair deformation increases.
This property can be explained by the BCS pair model.
This may indicate a prominent difference
in the effect of the deformation between
the spatial rotation and the pair rotation.
Further investigation is under progress.

\bigskip
\noindent{\bf\large Acknowledgments}\newline

This work is supported in part by JSPS KAKENHI Grant
Nos. 18H01209, 20K14458, 20K03964, 23H01167, 23K25864,
and by JST ERATO Grant No. JPMJER2304.
This research in part used computational resources provided 
by the Multidisciplinary Cooperative Research Program in the Center for
Computational Sciences, University of Tsukuba.

\bibliography{nuclear_physics,myself,current1,current2}

\end{document}